\documentclass[a4papper,11pt]{article}
\usepackage{graphicx}
\usepackage{epsf}%{epsfig}
\usepackage{amsmath}
\usepackage{amssymb}
\usepackage{wasysym}
\usepackage{enumitem}
\usepackage[font=small,labelfont=bf,labelsep=period,justification=centerlast]{caption}

\setlist{noitemsep, nolistsep}

\title{\bf Suggesting a very simple experiment designed to detect tachyons}
\author{\bf Jerzy Klemens Kowalczy\'nski}
\date{}

\begin{document}
\maketitle

    Retired from the Institute of Physics, Polish Academy of Sciences

    Correspondence address: J.K. Kowalczy\'nski, ON3, Institute of Physics, Polish Academy of Sciences, Al. Lotnik\'ow 32/46, 02-668 Warsaw, Poland

    E-mail: jkowal@ifpan.edu.pl

\begin{abstract}

Production of tachyons in, among other things, air showers would be in accordance with predictions of general relativity. Some such tachyons would travel with a precisely determined speed, almost equal to $5c/3$ relative to the earth, and would be registered high above the region of creation of air showers, e.g. on board of a satellite. A very simple experiment designed to detect these tachyons is outlined here. Brief justification to search for tachyons is also given.

\medskip


\end{abstract}

\section{Introduction}
In the present paper a very simple experiment designed to detect tachyons is outlined. The idea of the experiment stems from a hypothesis on the production of tachyons. This hypothesis is based on a realistic description (i.e. a realistic model) of the tachyonic phenomenon. This description is still only one such description to my knowledge, and it directly results from an exact solution of the Einstein-Maxwell equations.

The reader interested in the basis of the idea of the experiment can find it in my previous publications briefly presented in Sect. 2.

   In Sect. 3 the production of tachyons is schematically shown and described, which may be treated as a supplement to my previous publications.

The main topic of this paper, i.e. a scheme of a very simple experiment to search for tachyons moving with a speed almost equal to $5c/3$ ($c$ being the speed of light in vacuum), is presented in Sect. 4.

In Sect. 5 the search for tachyons is justified.

\section{Information on previous publications}
The geometric standards of recognition of the solution in question are given in [1], where the solution is presented by relations (1.2), (2.1), (2.3), and (3.2). Its physical interpretation is given in detail in book [2], where the solution is referred to as $\Omega_1$. In [2] also the hypothesis under consideration is presented and largely justified, and general ideas of the experiments resulting from this hypothesis are discussed.

   Book [2] contains many details, discussions, and laborious calculations; and therefore, to avoid deeper studies, the solution, premises of the hypothesis, and the hypothesis itself are given in an abbreviated form in [3], where comments on the empiric possibilities of tachyon production are limited to high energy collisions with atomic nuclei other than protons. In these comments many types of experiments to search for tachyons are discussed.

   If the reader prefers to get to know only the hypothesis and the comments on the empiric possibilities, he may make use of [4]. Section 4 in [3] and Sect. 2 in [4] (the hypothesis) are very similar. Section 5 in [3] and Sect. 3 in [4] (the comments) are almost identical. The ideas of experiments presented in the latter sections (i.e. in [3,4]) seem to be quite sufficient to perform such experiments. However, the reader who prefers to have these ideas aided by illustrations, will find the latter in this paper (as a supplement to [3,4]).

\section{Production of tachyons}

   The preliminary stage of the tachyon production is shown in Figs. 1 and 2. In the cases (A) and (B) of Fig. 2 peripheral nucleons are struck (``tangent'' collisions), which provides the most effective production conditions, and the electric type tachyon [2,3] (i.e. e-tachyon [3,4]) will immediately be produced. In the case (C) the proton present in the deuteron is struck and the magnetic type tachyon [2,3] (i.e. m-tachyon [3,4]) will immediately be produced. This tachyon (e- or m-) is called \textit{principal} [3,4]. It can be produced alone or together with its \textit{accompanying} tachyons [3,4] (see Figs. 1 and~3). The latter tachyons cannot be produced alone, i.e. without the principal tachyon [3,4]. It seems obvious that in the experiments performed in or at accelerators (high-energy colliders would be the best choice) only the cases (A) and (C) of Fig. 2 can take place, with electrons or antiprotons as striking particles. In the air showers initiated by cosmic (primary) particles of energy of    ${\sim} 10^{13}$--$10^{14}$ eV and higher, all the cases (A), (B), and (C) of Fig. 2 should be possible.

\begin{figure}[bht]%
  {\centering\small
  \includegraphics{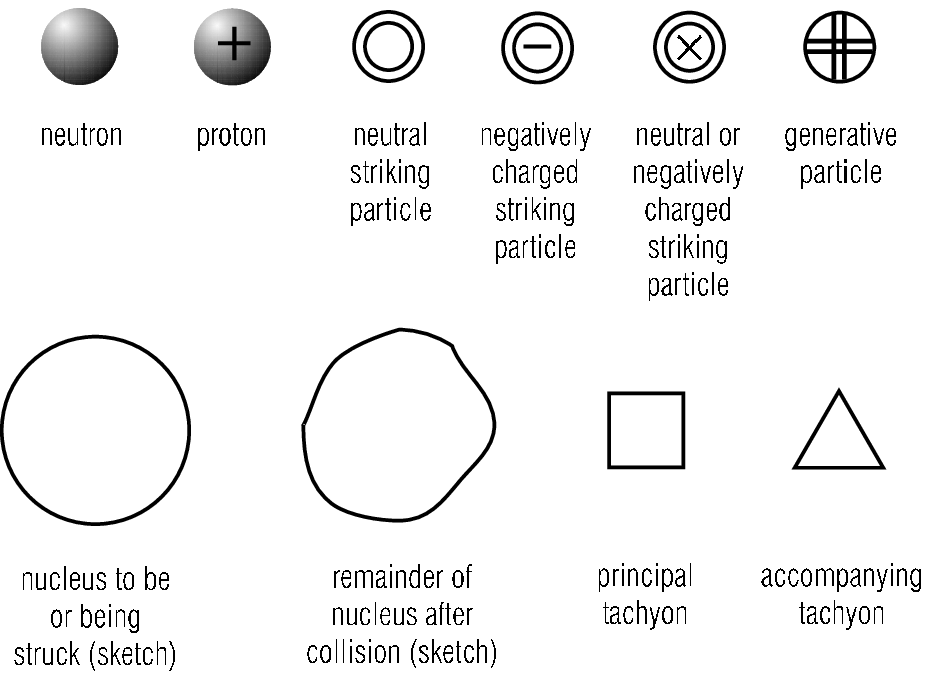}\\%[width=*]
  \caption{Objects shown in Figs. 2 and 3.
}\label{Fig1}}%
\end{figure}

\begin{figure}[thb]
  {\centering\small
  \includegraphics{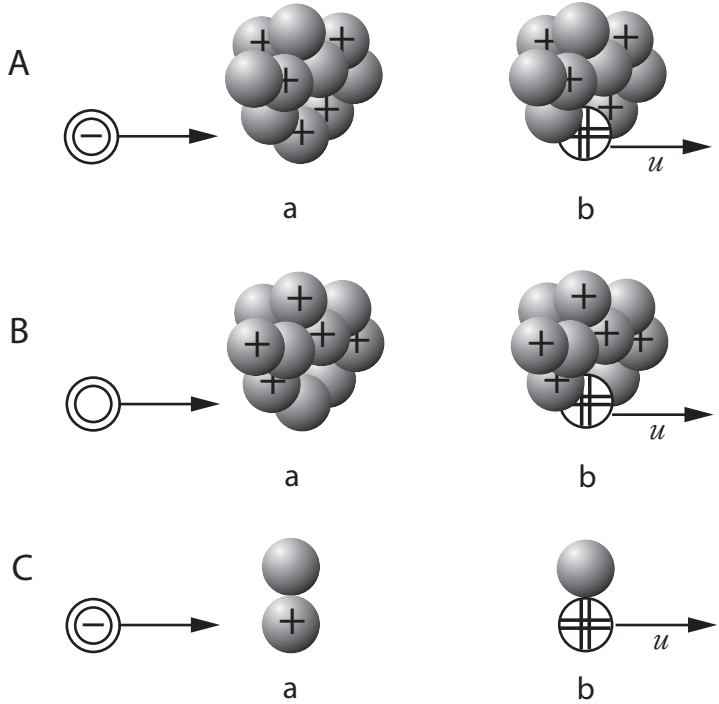}\\%[width=*]
  \caption{Examples of the preliminary stage of the tachyon production. The striking particle will collide with a nucleon of a nucleus (subcases (a)) and as a result of the collision a neutral particle (called the generative particle [2-4]) will be created. \textit{In statu nascendi} this particle moves with a velocity $u$ relative to the nucleus (subcases (b)). In the cases (A) and (C) the striking particle is negatively charged and it collides with a proton of the nuclei, and in the case (B) it is neutral and collides with a neutron of the nucleus. In the cases (A) and (B) e-tachyons and in the case (C) an m-tachyon will immediately be produced, if  $|u|$  is sufficiently close to $c$.
}\label{Fig2}}%
\end{figure}

According to our hypothesis the \textit{generative particle} [2--4] must be in a suitable electromagnetic field (in the \textit{initiating field} [2--4]) in order to be converted into a tachyon (principal, alone or together with its accompanying tachyons). This field is given by relations (11)--(13) and (19) in [3], or by relations (1)--(3) and (8) in [4]. Those relations describe the initiating fields in terms of the proper reference frame of the generative particle. Simple estimation shows that \textit{in statu nascendi} the generative particle must move with a velocity $u$, relative to the nucleus being struck, such that $|u|\cong c$ ($|u|<c$), in order to find itself in the initiating field demanded by our hypothesis (relativistic intensification of electromagnetic field; $u$ must not be confused with $v$ occurring in [3]). This means that $U\cong 1$ ($U>1$) since according to the relativistic formulae of transformation of electromagnetic field there is
\begin{equation}
    U=c/|u|,
\end{equation}
where $U$ is defined in [2--4]. Thus, according to our hypothesis, the principal tachyon produced has a velocity $w$ in the reference frame of the generative particle such that $|w|\cong c$ ($|w|>c$; cf. [3,4]) and $W$ (i.e. either $W_{+}$ or $W_{-}$) in the reference frame of the nucleus being struck (see Fig. 3; the terms \textit{forward} and \textit{backward} concern only the principal tachyons [3,4]). It should be emphasized that the condition   $U \cong 1$ means that all the velocities under consideration, including the velocity of the striking particle, have practically the same direction (cf. remarks and formulae concerning the angle~$\alpha$; given briefly in [3,4] and in detail in [2]). As regards the senses of these velocities, see Fig.~3. From the above mentioned relations describing the initiating fields we obtain [2--4] that
\begin{equation}
    U^2=5-4c^2/w^2.
\end{equation}

\begin{figure}[htb]
  {\centering\small
  \includegraphics{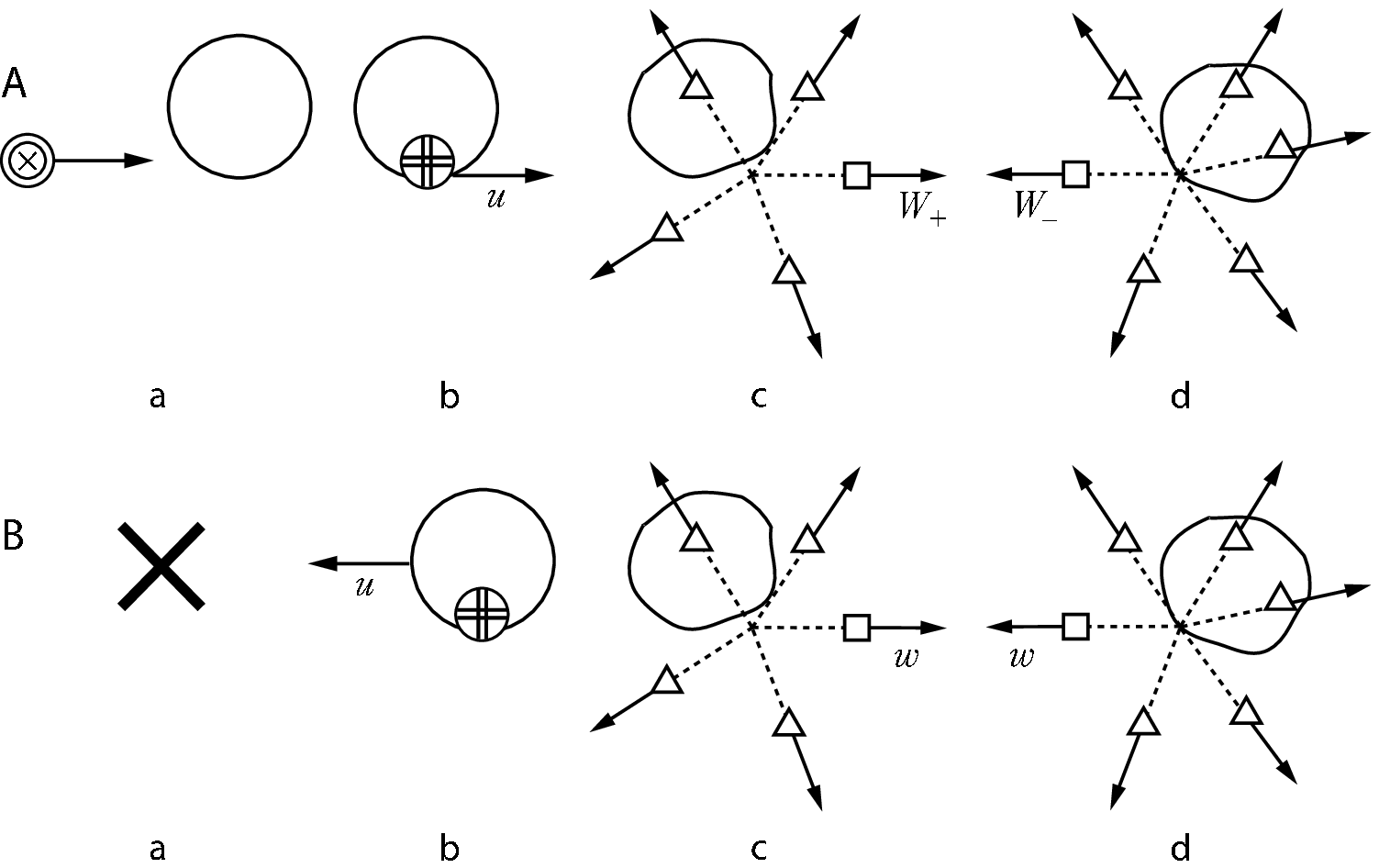}\\%[width=*]
  \caption{\looseness-1 Examples of the whole process of the tachyon production presented in two reference frames, i.e. in the proper reference frame of the nucleus to be (or being) struck (case (A)), and in the proper reference frame of the generative particle (case (B)). Subcases (b)--(d) show the same situation in these two reference frames respectively. Subcase (a) of the case (B) is absent since the generative particle does not yet exist. Subcases (a) and (b) of the case (A) are presented in a more detailed way in the subcases (a) and (b), respectively, in Fig.~2. The remainder of the nucleus being struck occurring in the subcases (c) and (d) is as follows: in the cases concerning the cases (A) and (B) in Fig.~2 it is either an excited nucleus or a group of separate smaller nuclei including, perhaps, subatomic particles, whereas in the cases concerning the case (C) in Fig.~2 it is that neutron only. In the subcases (c) the principal tachyon is the forward one, and in the subcases (d) the principal tachyon is the backward one. The principal tachyon and its accompanying tachyons (if exist) are born simultaneously at their common creation point. We have here: $|W_{+}|\cong c$ ($|W_{+}|>c$), $|W_{-}|\cong 5c/3$ ($|W_{-}|>5c/3$), $|w|\cong c$ ($|w|>c$).
}\label{Fig3}}%
\end{figure}

   In the present paper we are interested in the production of tachyons in air showers, and therefore we are interested in velocities of the principal tachyons in the reference frame of the nucleus to be struck (Fig. 3 (A)). Indeed, this frame is a terrestrial one, i.e. a very convenient laboratory reference frame, since thermal speeds of air molecules are negligible in comparison with $c$. According to the relativistic law of addition of velocities having the same direction, in this reference frame the principal tachyons have the velocity $W$ such that
\begin{equation}
    W=(u+w)/(1+uw/c^2)
\end{equation}
and from Eqs. (1)--(3) we obtain
\begin{equation}
    |W_{+}|=c[2U+(5-U^2)^{1/2}]/[U(5-U^2)^{1/2}+2]
\end{equation}
for the forward tachyon, and
\begin{equation}
    |W_{-}|=c[2U-(5-U^2)^{1/2}]/[U(5-U^2)^{1/2}-2]
\end{equation}
for the backward tachyon (see Fig. 3 (A)).

\goodbreak

From Eq. (4) we see that $|W_{+}|$ tends to $c$ when $U$ tends to unity. Thus the forward tachyons produced in air showers are very ``slow'' relative to the earth; so ``slow'' that indistinguishable as tachyons. In fact, if nuclei  $^{40}\!$Ar are struck so as to produce tachyons (e-tachyons in this case), then a rough estimation gives $|W_{+}|\apprle 1.0000008c=(1+8\times10^{-7})c$ and a value still smaller in the case of lighter nuclei. On the other hand, some tachyons accompanying these ``slow'' forward tachyons may move considerably faster, but then they escape from the showers sidewise and, as far as I know, the hitherto performed experiments to search for tachyons have not been designed to detect the tachyons escaping sidewise. All these factors can explain the failure of the just mentioned experiments. The formula determining the relations between the velocity of a principal tachyon and the velocity of its accompanying tachyon and the angle between these velocities is given in footnote 11 in [3] and in footnote 3 in [4].

It is easy to calculate from Eq. (5) that $|W_{-}|$ tends to $5c/3$ when $U$ tends to unity. Thus the backward tachyons produced in air showers are considerably faster than light in the laboratory (terrestrial) reference frame and therefore they can be easily identified as tachyons. What is more, $|W_{-}|\cong 5c/3$ ($|W_{-}|>5c/3$) in every case of such a production, i.e. there is practically only one precisely determined speed. In fact, if nuclei $^{40}\!$Ar are struck so as to produce tachyons (e-tachyons), then a rough estimation gives $|W_{-}|\apprle 1.67c = 1{.}002\times 5c/3$ and still smaller values in the case of lighter nuclei. This makes it possible to employ a very simple electronic system in the experiment discussed in the next section.

\section{The scheme of the apparatus designed to detect\newline backward tachyons and an idea of the experiment}

   The proposed apparatus is outlined in Fig. 4 and concerns an experiment of time-of-flight type. When a signal from D2 to ES follows that from D1 after time equal to $3l/5c$, where $l$ is the effective length of TD (i.e. the distance between D1 and D2), then we have a desired event, and only such events coming from TD should be registered by ES (delayed coincidences). Such events would mean flights of backward tachyons through TD. Note that here we may assume the tachyon speed as equal to $5c/3$ since a surplus of ${\sim} 0.2\%$ (see the end of Sect. 3), or even ${\sim}0.5\%$ to have a safety margin, would probably be below the time resolution of ES. The fact that we deal with only one precisely determined speed enables us to apply a relatively simple ES. The only difficulty is the necessity to have the apparatus at a sufficiently high altitude (i.e. above the region where creation of air showers begins to be significant) for a sufficiently long time. A satellite (including the International Space Station of course) seems to be a choice better than a balloon.

\begin{figure}[htb]
  {\centering\small
  \includegraphics{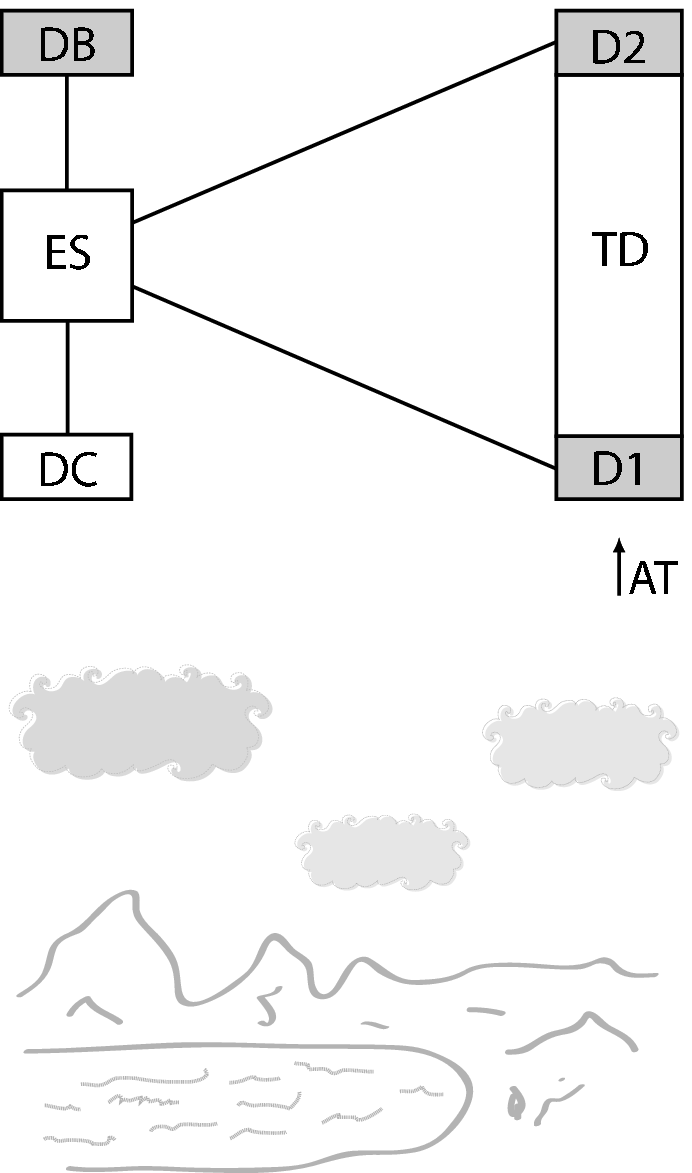}\\%[width=86mm]
  \caption{Apparatus designed to detect backward tachyons produced in air showers. AT -- arrival of tachyons. TD -- telescope of detectors of tachyons. D1 -- detector giving the first signal of tachyon's flight through TD. D2 -- detector giving the last signal of tachyon's flight through TD. DB -- detector of background. DC -- detector of clouds lying below the apparatus. ES -- electronic system.
}\label{Fig4}}%
\end{figure}

Detectors DB and DC are auxiliary. Data from DB enable us to calculate the probability that the desired events are apparent. The calculation would be simplified if D1, D2, and DB were identical. D1 and D2 may be common detectors of ionizing particles since according to the theory the tachyons under consideration are strongly ionizing objects.

The employment of DC (which should act night and day of course) is needed, and DC and TD must have the same visual field. Indeed, if it happens that the number of desired events above clouds is distinctly larger than that when DC does not see any cloud, or that the desired events take place only when DC sees clouds, then it is very probable that only the m-tachyon exists in nature (cf. footnote~1 in [4]). Note that the presence of deuterium is necessary to produce m-tachyons in air showers (see Fig. 2 (C); additional remarks are given in [3,4], and more details in [2]), and that the deuterium content is very low in the earth's atmosphere. The use of meteorological data instead of DC seems to be a worse choice.

\goodbreak

The telescope TD may contain intermediate detectors (Fig. 5 (b) and (c)) that may also be common detectors of ionizing particles. They would make the desired events more credible. The times of signals coming to ES from intermediate detectors need not be precisely determined by ES. Only a proper sequence of these signals (after that from D1 and before that from D2) would be sufficient (qualitative increase in the credibility of the desired events). Of course, by measuring times between all the detectors and obtaining the speed of $5c/3$ (even only approximately) everywhere, we would get quantitative (i.e. much stronger) increase in the credibility. This, however, needs ES much more complicated and/or TD considerably longer, which would increase the empiric difficulties. In both, i.e. qualitative and quantitative, cases the greater number of intermediate detectors the higher is the credibility of the experiment. This, however, is risky since we do not know the tachyon range in matter, i.e. in the material of the detectors in this case. Therefore a good idea is to connect independently two or more detector telescopes (all different) with ES. The presence of the simplest TD (Fig.~5 (a)) is then indispensable. Since the tachyon range in matter is unknown, D1's should not be shaded by, e.g., the satellite floor. Obviously, one of the possible solutions is to fasten TD's outside the satellite body.

\begin{figure}[htb]
  {\centering\small
  \includegraphics{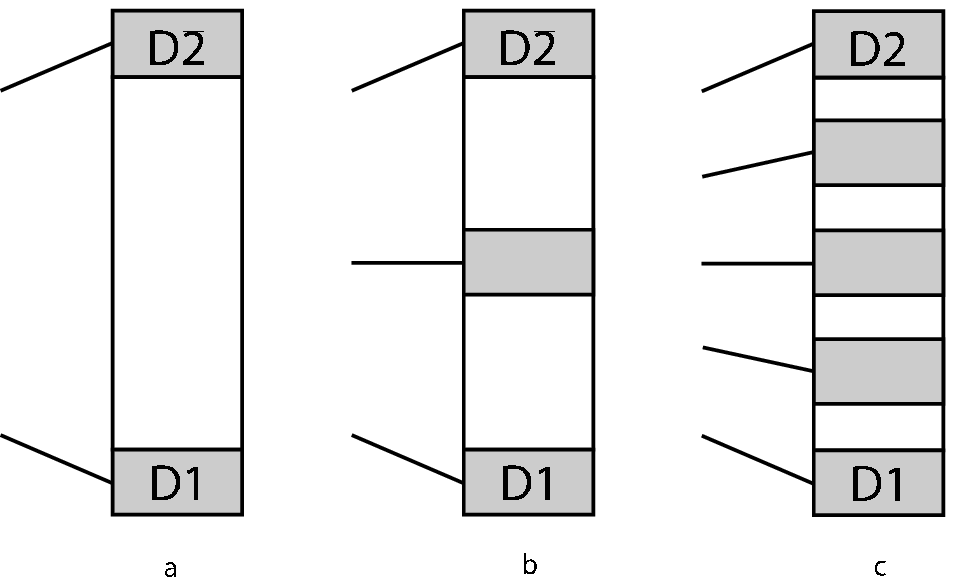}\\%[width=*]
  \caption{Various telescopes of detectors of tachyons: (a) the possible simplest telescope, (b) with one intermediate detector, (c) with several intermediate detectors.
}\label{Fig5}}%
\end{figure}

\section{Why the search for tachyons is rational}

   It is obviously tempting from the scientific point of view to overcome the light barrier and, in consequence, to observe or investigate some superluminal phenomena, as none of them has so far been empirically registered. This is indisputable. We may dispute, however, about the possible risk of wasting time, effort, and money, if the tachyon does not exist in nature. On the one hand, general relativity, being a serious theory, yields realistic descriptions of tachyonic phenomena; on the other, we do not know whether every realistic description given by this theory must relate to a situation in nature.

Surprisingly, all the above disputable problems also involve gravitational waves, as they have not yet been detected, and, what is more, general relativistic predictions of these waves and of tachyons result from such solutions of the Einstein equations that belong to the same family (of the Robinson-Trautman type). But in the case of gravitational waves the problem of risk is not taken into account. Very expensive and huge devices are used without positive results. Thus, in all fairness, the apparatus presented in Sect. 4 should be made and used, especially as it is cheap and simple, and therefore the above mentioned risk is not high, whereas the stake, i.e. overcoming the light barrier, is very high. Besides, being small and relatively light, this apparatus can only be an addition accompanying a main device on board of, e.g., a satellite. It should also be emphasized that the experiment outlined here is designed in accordance with general relativistic predictions, contrary to the hitherto performed experiments to search for tachyons.

Looking ahead, if gravitational waves and/or tachyons are detected, then we will register only slight signals from remote space in the case of these waves, whereas in the case of tachyons we may encounter, on the terrestrial scale, phenomena having surprising properties.

The above remarks may be summarized as follows: if the search for tachyons is thought to be irrational, then one should consider the search for gravitational waves just as, or even more, irrational.

\section*{References}

\begin{description}[leftmargin=\parindent,labelwidth=\parindent]
\item{[1]} J.K. Kowalczy\'nski, J. Math. Phys. \textbf{26} (1985) 1743
\item{[2]} J.K. Kowalczy\'nski, The Tachyon and its Fields, Polish Academy of Sciences, Warsaw (1996)
\item{[3]} J.K. Kowalczy\'nski, Acta Phys. Slovaca \textbf{50} (2000) 381, or hep-ph/9911441
\item{[4]} J.K. Kowalczy\'nski, Acta Phys. Polonica \textbf{B31} (2000) 523, or, with small extensions, hep-ex/0305008
\end{description}

\bigskip
\bigskip

\begin{appendix}

{\bf This article differs from its first version in the following:}
\begin{enumerate}
  \item Subscripts + and - at W's are enlarged in Fig.~3.
  \item A phrase in parentheses is inserted in the last sentence of the first paragraph of Sect.~4.
  \item In the last paragraph of Sect.~4 the phrase "... get quantitative (i.e. ..." replaces the misprint "...~~get~qualitative (i.e. ..." occurring in the first version.
  \item Two sentences are added at the end of Sect.~4.
  \item Figures 4 and 5 are slightly reduced in size.
  \item Instead of tilde marks, accent marks occur now over n's in the author name.
\end{enumerate}
\end{appendix}

\end{document}